\documentstyle[aps,twocolumn,prb,epsfig]{revtex}

\begin{document}
\draft

\title{On the Dynamics of the Evolution of HIV Infection}

\author{Rita Maria Zorzenon dos Santos \cite{rita}} 
\address{Instituto de F\'\i sica de S\~ao Carlos,
Universidade de S\~ao Paulo,  CP 369, CEP 13560-970, 
S\~ao Carlos, S\~ao Paulo, Brazil.}
\author{S\'ergio Coutinho }
\address{Laborat\'orio de F\'\i sica Te\'orica e Computacional,
Universidade Federal de Pernambuco, CEP 50670-901, Recife, Pernambuco,
Brazil.}

\maketitle

\begin{abstract} We use a cellular automata model to study the evolution
of HIV infection and the onset of AIDS. The model takes into account the
global features of the immune response to any pathogen, the fast mutation
rate of the HIV and a fair amount of spatial localization. Our results
reproduce quite well the three-phase pattern observed in T cell and virus
counts of infected patients, namely, the primary response, the clinical
latency period and the onset of AIDS. We have also found that the infected
cells may organize themselves into special spatial structures since the
primary infection, leading to a decrease on the concentration of
uninfected cells. Our results suggest that these cell aggregations, which
can be associated to syncytia, leads to
AIDS.
\end{abstract}

The human immunodeficiency virus (HIV), which causes AIDS (the acquired
immunodeficiency syndrome) has been the subject of most intense studies,
that
encompass diverse fields of scientific research, ranging from the natural
to the social sciences.  Major progress has been achieved by medical and
biological researchers in understanding the genetic code of the virus, the
role of its high mutation rate, the virus-host interaction, the apparent
failure of the immune system to control and eliminate the virus, and the
onset of AIDS. Nevertheless, the mechanisms by which HIV causes AIDS still
remain unexplained.

The immune response to any virus pathogen (virus, bacteria, etc.) is
generated by a complex web of interactions, involving the cooperative and
collective behavior of different types of white blood cells (monocytes, T-
and B-cells). The time scale involved to develop a specific immune
response may vary from days to weeks.  The complex dynamic of HIV
infection and the ensuing onset of AIDS involves a wide range of time
scales \cite{pantaleo93}. The primary infection exhibits the same
characteristics as any other viral infection: a dramatic increase of the
virus population during the first $2-6$ weeks, followed by a sharp
decline, due to the action of the immune system.  For HIV, however,
instead of being completely eliminated after the primary infection, a low
virus burden is detected for a long asymptomatic time: the clinical
latency period. This period may vary from two to ten (or more) years,
depending on the patient.  Besides the low concentration of HIV detected
 during this period, a gradual deterioration of the immune system is
manifested by the reduction of CD4$^{+}$T-cell populations in the
peripheral blood. The third phase of the disease is achieved when the
concentration of the T-cells is lower than a critical value ($\sim 30\%$),
leading to the development of AIDS. As a consequence, the patient normally
dies from opportunistic diseases, which would be usually controlled by the
immune system.  This common pattern observed during the course of the HIV
infection \cite{pantaleo93} is depicted in Fig. 1, which shows the plasma
viremia titer and the CD4$^{+}$T cell counts in the peripheral blood as
functions of time.

Several theories\cite{fauci88} have been proposed to explain why and how
the virus remain (albeit in low concentrations) in the organism after the
primary immune response, and the causes of the decline of T-cell counts,
leading to the onset of AIDS. So far, none of them has provided, a
complete explanation for the entire process. The most common assumption
made to explain the permanence of the virus is that of {\it fast mutation
rate of the HIV}, due to its rapid replication and its high degree of
variability \cite{nowak95,drosopoulos}. This assumption plays a central
role in our theory as well.

The challenge to understand the dynamics of the HIV infection has also
attracted the attention of mathematicians and physicists that have been
working on mathematical models \cite{perelson99} to describe different
aspects of the dynamics of the host-parasite interaction. Although many of
these models have contributed to the understanding of various aspects of
the development of the disease, they fail to describe either the changes
in the amount of virus detected in the body (blood, tissues and other body
fluids) during the entire process or the two time scales observed in the
course of the HIV infection, i.e., the primary response which runs from
days to weeks and the clinical latency period that may vary from
months to years. From the dynamical point of view, these different time
scales may be related to two distinct kinds of interactions: one local and
fast; and the other long-ranged and slow.

Most of the mathematical models proposed so far use
population equations, treating the infected patient as a homogenous
entity. Therefore local interactions and spatial inhomogeneities, caused
by localization of the initial of immune response in lymphoid organs, are
{\it not } taken into account. We believe that this feature, which is a
natural ingredient of our model, are of central importance.

Experimental evidences \cite{pantaleo97} support that the
lymphoid tissue is a major reservoir of HIV infection in vivo.
Moreover a snapshot of the distribution of cells among the different
compartments of the immune system will show only a small fraction
($2-4\%$) of the cells circulating in blood and lymph, while the majority
is found in the lymphoid organs \cite{steckel}.  Paradoxically, the
process of mobilization and activation of immune cells directed against
the virus by the immune system, which occurs in the lymphoid
micro-environment in this case, provides a milieu that contributes to the
virus spread.  \cite{pantaleo97}.

Such spatial structure emerges naturally in discrete models, based on
cellular automata that were shown~\cite{zorzenon99} to describe well
cooperative and collective patterns in experimentally observed immune
response \cite{zsb98}. Therefore we model the course of the HIV infection
by a cellular automaton. Our model takes into account the main features of
the immune response to any pathogen, the high mutation rate of the HIV and
a fair amount of spatial localization that may occur in the lymphoid
tissues. Our aim is to test whether the combination of these hypotheses
can explain the three-phase dynamics observed on the course of the HIV
infection (see Fig. 1).  The results obtained by simulations of our model
are shown in Fig. 2 and, as far as we know, this is the first time that
the complex dynamics of the HIV infection process has been so faithfully
reproduced by a theoretical model.

\begin{figure}
\leavevmode
\vbox{%
\epsfxsize=4cm
\epsfig{file=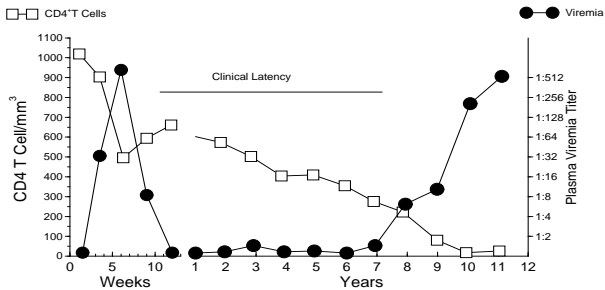,height=8cm,width=4cm,angle=-90}}
\caption{The common pattern exhibited by infected
patients, first presented by Pantaleo et al \cite{pantaleo93}. The plasma
viremia titer (black circles) and $CD4^+T$ cell counts (squares) versus
time show a three-phase dynamics.}
\label{fig1}
\end{figure}

We model the immune system cells in the lymphoid tissues by a
two-dimensional cellular automaton. To each lattice site we associate a
cell, that may be a target for the HIV (as for instance, a CD4+T cell or
a monocyte). Each cell can be in one of four states: (a)  {\it healthy};
(b) {\it infected-A1}, corresponding to an infected cell that is free to
spread the infection;  (c) {\it infected-A2}, an infected cell against
which a specific immune response has already developed and, finally (d)
{\it dead}, an infected cell that was  eliminated by the immune
response.

The initial configuration is mainly composed of healthy cells, with a
small fraction, $p_{HIV}$, of infected-A1 cells, representing the initial
contamination by the HIV.  In one time step the entire lattice is updated
in a synchronized parallel way, according to the rules described below.  
The updated state of a cell depends on the states of its eight nearest
neighbors.

\noindent {\it Rule 1: Update of a healthy cell:  (a): if it
has at least one infected-A1 neighbor, it becomes infected-A1. (b): if
it has no infected-A1 neighbor but does have at least $R$ ($2<R<8$)  
infected-A2 neighbors, it becomes infected-A1  (c): otherwise it stays
healthy.} 

\noindent Rule 1a mimics the spread of the HIV infection by contact,
before the immune system had developed its specific response against the
virus. Rule 1b represents the fact that infected-A2 cells may, before
dying, contaminate a healthy cell if their concentration is above some
threshold.

\noindent {\it Rule 2: An infected-A1 cell becomes  
infected-A2  after $\tau$ time steps.} 

\noindent An infected-A2 cell is one against which the immune response has
developed and hence its ability to spread the infection is reduced. Here
$\tau$ represents the time required for the immune system to develop a
specific response to kill an infected cell. Such a time delay is required
for each infected cell since in our model we view each new infected cell
as carrying a different lineage (strain) of the virus.  This is the way we
incorporate the fast mutation rate of the virus in our model. When a
healthy cell is infected, the virus uses the cell's DNA in order to
transcribe its RNA and replicate. During each transcription an error may
occur, producing, on the average, one mutation per generation and hence a
new strain of the virus is produced \cite{nowak95,drosopoulos}.

\noindent {\it Rule 3:  Infected-A2 cells become dead cells.} 

\noindent This rule simulates the depletion of the infected cells by the
immune response. 

\noindent {\it Rule 4: (a) Dead cells can be replaced by healthy cells
with probability $p_{ repl}$ in the next time step, (or remain dead with
probability $1-p_{ repl}$). (b) Each new healthy  cell introduced may be
replaced by an infected-A1 with probability $p_{infec}$.}

\noindent Rule 4a describes the bone marrow replenishment of the depleted
cells, mimicking the high ability of the immune system to recover from the
immunosupression generated by infection. As a consequence, it will also
mimic some diffusion of the cells in the tissue.  Rule 4b simulates the
entrance of new infected cells in the system that could come from other
compartments of the immune system.

This four-state automaton allows transitions from each of its four states
into the next one in a cyclic manner. We performed simulations of the
model on square lattices with periodic boundary conditions of $N=L^2$
sites, with $L$ ranging from $300$ up to $1000$.  All the parameters
adopted in the simulations were based on experimental data.  The initial
concentration of HIV, $p_{HIV}=0.05$, was chosen based on the observation
that the order of one in $10^2$ or $10^3$ T cells harbor viral DNA during
the primary infection \cite {schnittman}. We adopted $p_{infec}=10^{-5}$
due to
the fact that only one in $10^4$ to $10^5$ cells in the peripheral blood
of infected individual express viral proteins. Since the immune system has
a high ability to replenish the depleted cells, we used $p_{repl}=0.99$,
although smaller probabilities could also be considered, representing
different individuals. Since the time delay parameter ($\tau$)  may vary
from $2$ to $6$ weeks, we chose $\tau=4$. Each of our time steps
corresponds to one week.

We present in Fig. 2 the evolution of the densities of healthy cells,
infected cells (considering both A1 and A2 types) and dead cells, obtained
in simulations of our model . We show results averaged over one hundred
simulations (that used different initial configurations) and the
corresponding standard deviations (shown as error bars). There is
excellent qualitative agreement between our results for the density of
healthy and infected cells and the time evolution of the number of CD4+ T
cells in the peripheral blood and the plasma viremia titer showed in Fig.
1.  The model reproduces the two time scales observed in the dynamics of
the HIV infection:  a short one associated with the primary infection and
a long one related to the clinical latency period and the onset of AIDS.
The small error bars related to the first stage of the infection indicated
that its dynamics is
insensitive to the initial configuration. However, the large error bars
during the latency period suggest the presence of some mechanism
regulating this phase of infection.

\begin{figure}
\leavevmode
\vbox{%
\epsfxsize=8cm
\epsfig{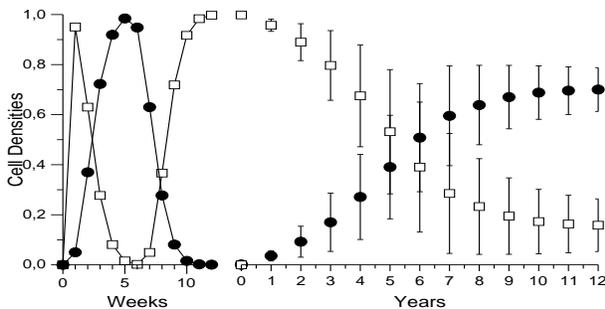}}
\caption{The results obtained from our simulations for a
two-dimensional lattice with $L=700$, $p_{HIV}=0.05$, $R=4$, $\tau=4$,
$p_{infec}=10^{-5}$, $p_{repl}=0.99$. The evolution of the population
densities exhibit the same three-phase dynamics observed for infected
patients. We have adopted {\it open squares} for healthy cells, 
{\it full circles} for infected cells.}
\label{fig2}
\end{figure}

The results presented in Fig. 2 are global properties, i.e. average
quantities over the entire system. Our model allows also a closer
look at {\it local} behavior, which in fact, may provide the
clue
for understanding also the global properties. From the analysis of
the spatial
configurations generated by the model in various individual simulations,
we noticed that the slow dynamics observed in the latency period is
related to the emergence, after completion of the primary response, of
some special spatial structures of infected cells. These growing special
structures spread the infection in such way that they slowly commit more
and more healthy cells, segregating and trapping uninfected cells.

In Fig. 3 we present four ``snapshots" of typical configurations obtained
during one particular simulation. Starting form an initial configuration
mostly composed of healthy cells (blue) with a random distribution of
infected-A1 cells (yellow), in subsequent time steps each individual
infected-A1 cell generates a pulse of infected cells, of width $(\tau+1)$, 
propagating in all directions. Whenever the average distance $\langle
l_{1}\rangle$ between individual infected cells in the initial
configuration is less than or equal $(2\tau +1)$, the independent pulses
achieve a maximum coverage of the lattice. For $\tau=4$ the maximum
coverage occurs after five weeks as shown in Fig. 3a. Note that the
distribution of dead (red) cells corresponds to the initial configuration
of infected cells. After that the concentration of infected cells
decreases to a minimal value at $2(\tau+1)$ time steps, establishing the
end of the primary infection phase ($10$ weeks, in this case).

\begin{figure}
\leavevmode
\vbox{%
\epsfxsize=6 cm
\epsfig{file=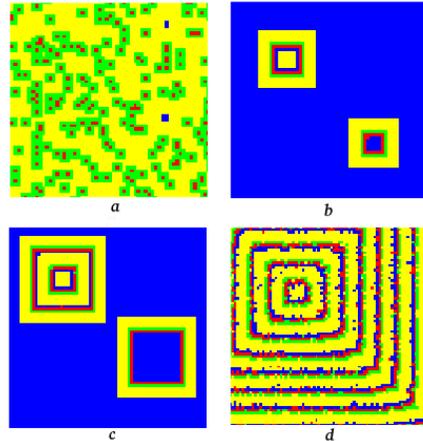,height=6cm,width=6cm}}
\caption{(Color) Four snapshots of parts of the lattice configuration
for different time steps:(a)-(d) correspond to $5$, $18$, $25$ and $200$
weeks, respectively. During the latency period we observe an
organization of different types of cells in layers which interpenetrate
each other. We have adopted the same parameters used in Figure 2.}
\label{fig3}
\end{figure}
In the following time steps the presence of infected cells will be
dictated by $p_{infec}$, according to rule 4b. When new infected cells
are introduced, they may generate two different kinds of structures. The
simplest case corresponds, as discussed above, to a wave of infected cells
propagating in all directions, but in this case since $p_{infec}<<p_{HIV}$
the average distance $\langle l_{2}\rangle$ between the new infected
cells, introduced with probability $p_{infec}$ at random locations, is
much larger than $(2 \tau+1)$. Examples of such structure are shown in the
bottom right part of Figs. 3b and 3c. The spread of infection generated by
these structures alone takes a longer time to reach nearly complete
lattice coverage, and after that the infection vanishes. It is very rare,
however, to find only these simple structures. A second type of structure
also occurs, due to an interplay between rules 1b, 4a and 4b.  These
special structures are generated by "sources" of infected cells. These
sources appear, for instance, when a new infected-A1 cell, introduced by
 rules 4a and 4b, is surrounded by at least four dead cells. At every
every ($\tau+3$) time steps they launch a propagating wave front of
infected cells with width ($\tau+1$). The period of the wave fronts
corresponds to the period of time the infected-A1 cell (source)  remains
in the same state ($\tau$) plus the three time steps to transition to
types infected-A2, dead and healthy, in this sequence. Figs 3b and 3c
(upper left) show the growing of such structures for two subsequent
periods, corresponding respectively to $18$ and $25$ weeks. As these
structures grow, the number of infected cells increases and the
concentration of healthy cells decreases. Moreover they segregate and trap
uninfected cells between two consecutive waves of infected cells.  These
special pattern of cell aggregations explain the slow decrease of T cells
in infected patients during the latency period. These growing structures
may eventually cover the entire lattice, with final densities of different
cells evolving towards a steady state with average concentrations:
$(\tau+1)/(\tau +3)$ for all the infected cells, and $1/(\tau+3)$ for both
the dead and healthy cells. Note that the average density for healthy
cells at the steady state is always lower than the observed critical
value($\sim 30\%$) of CD4$_{+}$ T cells counts related to the onset of
AIDS in infected patients, i.e, the system's breakdown. Therefore the
dynamics of real experimental data is related to the transient behavior of
our model and not to its steady state. The long time scale, associated
with this transient period that follows the primary infection, during
which the model relaxes towards its steady state, corresponds to the
clinical latency period.

Analysis of the source distribution at any given time shows that the
latency period depends on $\langle l_{3}\rangle$, the average distance
between sources and, consequently, on the probability of occurrence of
sources ($\sim p_{repl}.p_{infec}.p_{HIV}$). Actually, since new sources
can be released at any time step, the length of this transient time
depends on the spatio-temporal average of the distance between sources.

In this work we have shown that our cellular automaton model reproduces
quite well the three-stage dynamics observed in the course of the HIV
infection, and the different time scales observed in this process. The
short time scale, characteristic of the primary infection, increases when
$\tau$ is increased or when $p_{HIV}$ decreases. The long time scale,
responsible for the clinical latency period and the onset of AIDS, is
associated with the emergence of special structures, that increase slowly
the number of infected cells and confine healthy cells. This special
pattern formation depends on the value of the parameter $R$ of rule 1b,
$p_{repl}$ and $p_{infec}$.  Smaller values of $p_{repl}$ enhance the
presence of dead cells and favor the formation of the special growing
structures. We have also performed a careful investigation of the
parameter space, and found that three-phase dynamics is reproducible for a
wide range of the parameters. The complete study of the parameter space,
the detalied discussion of the necessary conditions to generate the
special spatial structures and the role they play spreading the infection,
is under preparation and will be published elsewhere.

The growing structures of infected cells may be associated with syncytia,
the aggregation of cells observed experimentally, suggesting
according to our results that they are responsible for the depletion of T
cells leading to AIDS. These results actually corroborate some previous
suggestions that syncytia are responsible for the permanence of the virus
in the system, based on the analysis of the similarities between HIV
infection and other diseases \cite{fauci88,weiss}.

Finally we emphasize that our cellular automaton model is, as far as we
know, the first model which succeeded to reproduce the general features
of the complex dynamics of the HIV infection, as observed in infected
patients.  This work further substantiates claims made in previous studies
\cite{zsb98}, that discrete models may be useful to describe emergent
properties of complex biological systems, and to understand the mechanisms
underlying its dynamical behavior. The reason for our success in
describing the three-stage dynamics, whereas the other approaches fail, is
that they do not take into account the spatial localization effects that
play a major role on the course of the infection. A detailed study of the
mechanisms underlying the dynamics described by our model may lead to new
differential equation approaches, more suitable to describe the kind of
dynamics observed in the course of HIV infection.

We thank Dr. M. Curi\'e Cabral, I. Procaccia, S. Bocalletti, and J-P.
Eckmann for enlightening discussions and to Borko St\v{o}sic for
assistance with color graphics.  RMZS' thanks J.F. Fontanari for the
hospitality during her visit to IFSC-USP, and FAPESP (project 99/09999-1)
for supporting her stay. This work was partially supported by the
Brazilian Agencies CNPq, CAPES, FINEP (under the grant PRONEX
94.76.0004/97) and FAPERJ.\\



\end{document}